\documentclass[%
 reprint,
 amsmath,amssymb,
 aps,
]{revtex4-2}
\newcommand\thefont{\expandafter\string\the\font}

\usepackage{graphicx}
\usepackage{dcolumn}
\usepackage{bm}

\usepackage{float} 

\usepackage{xcolor}
\usepackage{epsfig}
\usepackage{epstopdf}
\usepackage{glossaries}  
\usepackage{algpseudocode}
\usepackage{algorithm} 
\usepackage{siunitx}
\usepackage{comment}

\newacronym{abc}{ABC}{approximate Bayesian computation}
\newacronym{qed}{QED}{quantum electrodynamics}
\newacronym{sfqed}{SFQED}{strong-field quantum electrodynamics}
\newacronym{ml}{ML}{machine learning}
\newacronym{NN}{NN}{neural network}

\newcommand{\Cedit}[1]{{\color{cyan}#1}}

\begin{document}

\preprint{APS/123-QED}

\title{Prospects for statistical tests of strong-field quantum electrodynamics with high-intensity lasers}


\author{C.~Olofsson}
   \email{christoffer.olofsson@physics.gu.se}
\author{A.~Gonoskov}%
 \email{arkady.gonoskov@physics.gu.se}
\affiliation{%
Department of Physics, University of Gothenburg, SE-41296 Gothenburg, Sweden
}%



\date{\today}

\begin{abstract}

Exploiting high-energy electron beams colliding into high-intensity laser pulses brings an opportunity to reach high values of the dimensionless rest-frame acceleration $\chi$ and thereby invoke processes described by strong-field quantum electrodynamics (SFQED). 
Measuring deviations from the results of perturbative SFQED at high $\chi$ can be valuable for testing the existing predictions, as well as for guiding further theoretical developments. Nevertheless such experimental measurements are challenging due to the probabilistic nature of the interaction processes, a strong background produced by low-$\chi$ interactions and limited capabilities to control and measure the alignment and synchronization in such collision experiments. Here we elaborate a methodology of using approximate Bayesian computations (ABC) for retrieving statistically justified inferences based on the results of many repeated experiments even in case of partially unknown collision parameters that vary from experiment to experiment. As a proof of principles, we consider the problem of inferring the effective mass change due to coupling with strong-field environment.
\end{abstract}


\maketitle


\section{\label{sec:Introduction} Introduction}

Although fundamental principles of quantum electrodynamics (QED) are known for their precise experimental validations, the implications they purport for sufficiently strong electromagnetic fields remain theoretically intricate and lack experimental data. Colliding accelerated electrons with high-intensity laser pulses can be seen as a newly emerging pathway to such experimental data \cite{poder2018experimental,cole2018experimental,abramowicz2021conceptual,yakimenko2019facet}. The local interaction is characterized by the dimensionless ratio of the electron acceleration in its rest frame to the acceleration that would be caused by the Schwinger field $E_\text{crit}$:

\begin{equation} \label{eq:chi}
\chi = \frac{\gamma_e}{E_\text{crit}} \sqrt{\left(\vec{E} + \left(\vec{v}/c\right) \times \vec{B}\right)^2 - \left(\vec{E} \cdot \vec{v} /c\right)^2}
\end{equation} 
where $\vec{v}, \gamma_e$ are the velocity and Lorentz factor of the electron, whereas $\vec{E},\vec{B}$ are the electromagnetic field vectors. Here, $E_\text{crit}=m_e^2 c^3/q_e\hbar\approx 10^{18} ~ \si{\volt\per\meter}$ where $\hbar$ is the reduced Planck constant, $c$ is the speed of light and $m_e, q_e$ are the mass and charge of the electron respectively. At $\chi \ll 1$ the electrons are subject to classical emission and corresponding radiation reaction. Emission of photons and corresponding recoils at $\chi \sim 1$ are described by non-linear Compton scattering and have been experimentally observed in several experiments \cite{bula1996observation,iinuma2005observation,kumita2006observation,englert1983second}. Measuring quantitative properties of the photon emission (e.g. energy, angular or polarization distribution) at $\chi \sim 1$ can be perceived as a logical next step, while results for $\chi \gg 1$ can potentially facilitate theoretical developments or even lead to fundamental discoveries (see Ref. \cite{fedotov2022advances} and references therein). \\

A severe obstacle for the outlined efforts is the interaction complexity. The value of $\chi$ for each electron in the beam varies in time and overall depends on the electron position relative to the laser pulse location, which can also vary from experiment to experiment due to spatio-temporal mismatches. For contemporary laser pulse durations, many electrons can lose a significant part of their initial energy prior to reaching the strong-field region, where they have a chance to emit at high $\chi$. Additionally, due to the Breit-Wheeler process the emitted photons can decay into electron-positron pairs, which can lead to the onset of an electromagnetic cascade. In combination, this means that the measurable post-collision distributions of photons, electrons and positrons are predominantly determined by low-$\chi$ emissions, giving no direct information about emissions at high-$\chi$, even if they had been invoked. \\

One known way of dealing with such difficulties is Bayesian binary hypothesis testing, which is based on comparing experimental results with the outcomes computed on the basis of each of two competing theories. However, even in the absence of a distinct hypothesis to be tested, one can use a similar technique to determine parameters that quantify deviations from the approximate theory (sometimes referred to as parameter calibration procedure \cite{ritto2022reinforcement,kennedy2001bayesian,dejong1996bayesian}), which in our case can be the theory on non-linear Compton scattering valid for moderate $\chi$ values. One practicality of this approach is the possibility to gain statistically rigorous knowledge from many experiments even in case of low repeatability. For example, the inference about high $\chi$ events is feasible regardless if the alignment of the laser-beam setup varies uncontrollably between experiments which we cannot measure. \\

In this paper we consider the possibility of using the technique of approximate Bayesian computation (ABC) in the forthcoming experiments \cite{brehmer2020mining,akeret2015approximate,ritto2022reinforcement}. As a proof-of-principle problem we elaborate the use of this method for measuring the constant that quantifies the effective mass shift \cite{fedotov2022advances,yakimenko2019prospect,ritus1970radiative,meuren2011quantum}. We assess the use of the ABC technique in the context of possible experimental conditions and analyze main requirements, difficulties and opportunities for improvements. The paper is arranged as follows. In Sec. \ref{sec:methodology} we motivate the use of likelihood-free inference and state the ABC algorithm. In Sec. \ref{sec:problemstatement} we demonstrate a proof-of-principle approach to infer the effective mass change, assessing the difficulties and limitations. Sec. \ref{sec:analysis} provides the numerical aspects in simulating the experiment and gives the prospects of the outlined methodology. We make conclusions in Sec. \ref{sec:Conclusion}.

\section{Methodology} \label{sec:methodology} 
Before turning to the subject-specific analysis, let us consider the methodology using a general problem formulation. Suppose we study a probabilistic process by carrying out experiments. Each experiment yields measurement data $x_{\text{obs}}$. We have a model $M(\theta, z)$ that gives predictions $x = M(\theta, z)$ for this data for any given value of a model parameter $\theta$ and a \textit{latent} parameter $z$. Here $\theta$ is a fundamental parameter that quantifies the process itself and thus its unique value is of interest, whereas $z$ denotes an unmeasured parameter that can vary from experiment to experiment and determines the outcome $x$ in accordance with model $M$. We assume that there exist a value of $\theta$ for which the model describes (to some extent) observations given an appropriate value of $z$ for each experiment. Our task is to infer the probability distribution for the value of $\theta$ from a series of repeated experimental measurements. Put differently, the objective is to infer the most probable range for $\theta$ given the observed data $x_{\text{obs}}$. Bayesian statistics provides a framework for the outlined problem. The probability distribution to be determined is referred to as a posterior distribution $p(\theta|x_{\text{obs}})$, which explicitly indicates the data $x_{\text{obs}}$ used for making the inference. Let us start from the case of no latent parameter. The posterior can then be calculated using Bayes' theorem

\begin{equation} \label{eq:BayesPosterior}
    p(\theta | x_{\text{obs}}) = \frac{p(x_{\text{obs}}|\theta) \cdot p(\theta)}{p(x_{\text{obs}})}
\end{equation}
where $p(\theta)$ quantifies the prior knowledge about possible values of $\theta$, the likelihood $p(x_{\text{obs}}|\theta)$ conveys how likely a measurement yielding $x_{\text{obs}}$ is for a given $\theta$ and $p(x_{\text{obs}}) = \int p(x_{\text{obs}}|\theta) p(\theta) d\theta$ appears as a normalizing factor. To incorporate the dependence on the latent parameter we integrate over all its possible values, denoting $p(x_{\text{obs}}|\theta, z)$ as the corresponding joint likelihood
\begin{equation} \label{eq:BayesPosteriorZ}
    p(\theta | x_{\text{obs}}) = \frac{ \int p(x_{\text{obs}}|\theta, z) p(z) dz \cdot p(\theta)}{\iint p(x_{\text{obs}}|\theta, z) p(z) p(\theta) dz d\theta},
\end{equation}
where $p(z)$ specifies prior knowledge related to values of the latent parameter $z$. Now we can sequentially account for all observations, each time using the obtained posterior as the prior for processing the next observation. Note that we do not update the prior for $z$ because its value is assumed to be different in all the experiments. \\

A closed form of the posterior rarely exist and numerical approaches are often used. A common strategy is to approximate the posterior by collecting a finite number of samples from it. Methods such as importance sampling, Markov chain Monte Carlo (MCMC) and sequential Monte Carlo (SMC) \cite{tokdar2010importance,doucet2001introduction,brooks2011handbook} are prevalent choices. However, all of the above will require direct evaluation of the likelihood which can be computationally prohibitive for highly dimensional datasets \cite{sisson2018handbook}. If the model $M$ is implicitly defined through a computer simulation, its concomitant likelihood can be intractable \cite{brehmer2020mining}. A remedy is offered by the rapidly developing field of simulation-based inference \cite{cranmer2020frontier} in which the direct calculation of the likelihood is averted. To motivate its use we adopt and develop the discussion made in Ref. \cite{sisson2018handbook}. \\ 

Consider the standard rejection sampling algorithm with the goal of sampling a target density $T(\theta)$ provided some auxiliary sampling density $A(\theta)$ with the requirement $A(\theta) > 0 \  \mathrm{if} \  T(\theta) > 0$. Then, the algorithm reads
\begin{algorithm}[H] 
\caption{Standard rejection sampling algorithm}
\label{alg:StandardRejectionSampling} 
\begin{algorithmic}[1]
\State Sample a proposal $\theta^* \sim A(\theta)$.
\State Admit the proposal with a probability of $\frac{T(\theta^*)}{C A(\theta^*)}$ where $C \geq \mathrm{argmax[\frac{T(\theta)}{A(\theta)}]}$.
\State If $\theta^*$ was not admitted, discard the proposal and repeat steps (1)-(2) as many times necessary.
\end{algorithmic}
\end{algorithm}
After $N$ trials a collection of samples from $T(\theta)$ is obtained. The connection to Bayesian statistics is made by selecting
$T(\theta) = p(\theta | x_{\text{obs}})$ and $A(\theta)=p(\theta)$. Then, Eq. \eqref{eq:BayesPosterior} implies that the acceptance rate in Alg. \ref{alg:StandardRejectionSampling} becomes proportional to the likelihood $\frac{p(\theta^* \lvert x_{\text{obs}})}{p(\theta^*)} \propto p(x_{\text{obs}} | \theta^*)$ which is incalculable by our premise. Still, it is possible to determine whether to accept proposals or not without explicit computation of the likelihood. To show this we first note that the model $M(\theta,z)$ is capable of generating samples of observations $x \sim p(x_{\text{obs}}|\theta, z)$ provided values of $\theta$ and $z$. Now, the probability to produce $x = x_{\text{obs}}$ coincides with $p(x_{\text{obs}}|\theta, z)$ which calls for modifications to Alg. \ref{alg:StandardRejectionSampling} so that it reads

\begin{algorithm}[H]
\caption{: Likelihood-free rejection sampling} \label{alg:LikelihoodRejectionSampling}
\begin{algorithmic}[1]
\State Sample proposals $\theta^* \sim p(\theta)$, $z^* \sim p(z)$.
\State Generate data $x^* = M(\theta^*, z^*)$ from the model.
\State If $x^* = x_{\text{obs}}$ the proposal is admitted, if not it is discarded.
\State Repeat (1)-(3) as many time necessary.
\end{algorithmic}
\end{algorithm}
While avoiding direct computation of the Likelihood, step $3$ of Alg. \ref{alg:LikelihoodRejectionSampling} introduces a notable impediment.
To illustrate it, consider the binning of data from an experiment into $\text{dim}(x_{\text{obs}})=B$ bins so that 

\begin{align}
    x_{\text{obs}} &= \left[c_1, c_2, c_3, ..., c_{B}\right], \label{eq:DobsEx} \\
    x &= \left[c'_1, c'_2, c'_3, ..., c'_{B}\right] \label{eq:DsimEx}
\end{align}
where $c_b, c'_b \in \mathbb{Z}$ denote integer counts belonging to the $b$:th bin. Then, denote $p_b$ as the probability to coincide $c_b = c'_b$ at bin $b$, assuming that this is independent between bins. Then, the probability to accept a proposal $\theta^*$ becomes

\begin{equation} \label{eq:ExactData}
    p(x = x_{\text{obs}}) = \prod_{b=1}^{b=B} p_b
\end{equation}
which approaches zero in the limit of highly dimensional datasets $B \to \infty$. The acceptance rate in Eq. \eqref{eq:ExactData} is lower or even infeasible for continuous data in which $c_b, c'_b \in \mathbb{R}$ are real numbers. Hence, the appeal for a precise match has to be relieved in making the sampling efficiency practical. Realizing that this rate becomes significantly higher by admitting samples if $x \approx x_{\text{obs}}$ prompts us to define a rule when data are \textit{sufficiently close}

\begin{equation} \label{eq:AlliviateCondition}
    ||x - x_{\text{obs}}|| \leq \epsilon 
\end{equation}
where $||\cdot||$ is a suitable distance metric and $\epsilon$ is a threshold. Accepted samples in accordance with Eq. \eqref{eq:AlliviateCondition} are inevitably drawn from an approximate posterior $\hat{p}(\theta|x_{\text{obs}})$ and its accuracy is solely dictated by $\epsilon$ which also affect the sampling efficiency. However, consider the aforementioned example with an Euclidean distance metric so that Eq. \eqref{eq:AlliviateCondition} reads

\begin{equation} \label{eq:EuclideanMetric}
    \left( \sum_{b=1}^{b=B} \left(c_b - c'_b\right)^2 \right)^{1/2} \leq \epsilon
\end{equation}
and examine the favorable case in which $c_b - c'_b \sim \Delta \ll 1$ varies negligibly between bins. We can then naively state Eq. \eqref{eq:EuclideanMetric} as

\begin{equation} \label{eq:BBounded}
    \text{dim}(x_{\text{obs}}) \leq \left( \epsilon/\Delta \right)^2 .
\end{equation}
Evidently, Eq. \eqref{eq:BBounded} states that the dimension of $x_{\text{obs}}$ is bounded from above by the threshold $\epsilon$ and the error $\Delta$. However, for the quality of inference $\epsilon \to 0$ is desired, which puts a stringent limit on the dimensionality of $x_{\text{obs}}$. To mitigate this, one can introduce so-called \textit{summary statistics}
\begin{equation} \label{eq:SummStat}
    S : \mathbb{R}^B \mapsto \mathbb{R}^{\beta}
\end{equation}
being a function that transforms data of potentially noisy nature into a vector of indicative characteristics ought to unambiguously characterize the data with respect to all possible $\theta$. Clearly, the dimensionality $\beta$ of the space of such vectors can be much less than the number of cells $B$. Moreover, the function of summary statistics can even be defined in an agnostic way with respect to the binning choice. As an example, one could construct a vector containing the sample mean $\mu$ and variance $\sigma^2$ of $x_{\text{obs}}$: $S(x_{\text{obs}}) = \left(\mu, \sigma^2 \right)$. \\

By converting $x_{\text{obs}} \to S(x_{\text{obs}})$, the third step of Alg. \ref{alg:LikelihoodRejectionSampling} can be reformulated to accept samples if
\begin{equation} \label{eq:SummStatReject}
    ||S(x) - S(x_{\text{obs}})|| \leq \epsilon.
\end{equation}
Although we now have methodologically accurate and in some cases practically feasible routine for sampling the posterior there are two more standard improvements. First, Eq. \eqref{eq:SummStatReject} implies an acceptance probability of either zero or one and thus doesn't account for how close the match is. To enhance the contribution of the cases yielding more accurate agreement relative to the ones giving a marginal agreement, one can use a so-called \textit{kernel function}
\begin{equation} \label{eq:kernel}
    K_{\epsilon} : \mathbb{R}^{\beta} \mapsto \mathbb{R} = K_{\epsilon} \left( \frac{ || S(x) - S(x_{\text{obs}}) ||}{\epsilon} \right),
\end{equation}
which defines a probability transition from one in case of a perfect match ($K_{\epsilon}(0) = 1$) to zero in cases of deviation by the summary-statistics distance of order $\epsilon$ and greater. \\

The second improvement concerns the fact that Alg.~\ref{alg:StandardRejectionSampling} implies either accepting or rejecting cases, which means that many accepted cases are needed to mitigate the noise related to this additional probabilistic element in the algorithm. Effectively this means that we marginally benefit from cases of low acceptance probability. To avoid this, one can instead interpret the acceptance probability as the weight of samples, thereby accounting for all the proposals that yield non-zero acceptance probability. \\

We can now return back to the inclusion of the latent variable $z$. In this case, we can generate several proposals $z^* \sim p(z)$ based on our prior knowledge of it and again accept the cases of good enough matches based on the outlined procedure. Effectively, we try to guess $z$ using as many attempts as needed. Finally, we note that we can sequentially update our posterior using each $x_\text{obs}$ in a sequence of measurements. To do so, we can compute the posterior for each new measurement using the previous posterior as the prior. The algorithm for processing the $i$-th observation ($i=1$ denote the first measurement in the sequence) $x^i_\text{obs}$ for computing the posterior $p\left(\theta \mid x^i_\text{obs}, x^{i-1}_\text{obs}, ..., x^1_\text{obs}\right)$ from the previous $p\left(\theta \mid x^{i-1}_\text{obs}, ..., x^1_\text{obs}\right)$ then takes the form

\begin{algorithm}[H]
\begin{algorithmic}[1]
\caption{: ABC sampling with latent variable}\label{alg:ABCsampling} 
\State Sample proposals $\theta^* \sim p\left(\theta \mid x^{i-1}_\text{obs}, ..., x^1_\text{obs}\right)$, $z^* \sim p(z)$.
\State Perform a simulation and retrieve $x^* = M(\theta^*$, $z^*)$ and compute the weight:
\begin{equation}
w^* = \frac{K_\epsilon\left(\| S(x^i_\text{obs}) - S(x^*)\|/\epsilon\right)}{p\left(\theta^* \mid x^{i-1}_\text{obs}, ..., x^1_\text{obs}\right) p(z^*)}
\end{equation}
\State If $w^* > 0$, accept the proposal with the computed weight. 
\State Repeat steps (1) -- (3) as many times as needed to approximate the posterior $p\left(\theta \mid x^i_\text{obs}, ..., x^1_\text{obs}\right)$.
\end{algorithmic}
\end{algorithm}
In practice, one central difficulty of the ABC routine is choosing valid summary statistics, i.e. summary statistics that differentiate all the cases in terms of $\theta$ and $z$. This means that summary statistics doesn't yield close states for any two different pairs of $\theta$ and $z$. Clearly, if this is not the case the procedure admits the acceptance of cases of wrong $\theta^*$ when $z^*$ provides a compensation to make $S(x^*(\theta^*, z^*))$ close to $S(x_\text{obs}(\theta^{true}, z^{true}))$. This can totally preclude the convergence of the ABC sampling procedure. Finding robust summary statistics is known to be a problem-dependent task that requires analysis of possible cases. In the next section we consider a proof-of-principle problem that includes a dependency on the latent variable. In doing so, we determine valid summary statistics and elaborate possible experimental strategies relevant to the tests of SFQED based on the collision of electron beams with focused laser pulses.

\section{Problem statement} \label{sec:problemstatement} 
As a proof-of-principle case, we consider the problem of detecting and measuring the extent of effective mass shift for the electron due to its coupling with the strong-field environment \cite{fedotov2022advances,yakimenko2019prospect,ritus1970radiative,meuren2011quantum}. The task is to infer the value of the parameter that quantifies this effect from the measured angular-energy spectra of photons emitted during the collision of high-energy electron beams with focused laser pulses. We make several assumptions to simplify the problem while keeping some indicative difficulties that show the capabilities of the methodology in question. In particular, we assume that the spatio-temporal mismatches between the electron beam and focused laser field are not measurable and vary from collision to collision. This leads to fluctuations of the electromagnetic field amplitude observed by the electrons. This in turn makes it impossible to relate the change of electron dynamics in a particular experiment (collision) to any certain amplitude, which has to be determined in the case of a straightforward measurement of the effective mass shift. To show how the ABC methodology resolves this difficulty we model the aforementioned variations by assuming that the electron beam propagates through a 1D laser pulse with an unknown amplitude that varies from collision to collision. In terms of introduced terminology, we introduce a latent parameter being a factor $< 1$ that reduces the laser field amplitude everywhere in each experiment, but varies uncontrollably from experiment to experiment. In what follows, we detail this \textit{model} of hypothetical experiments. \\

The presence of a strong background electromagnetic field is conjectured to drive the expansion parameter of QED to $\alpha_f \chi^{2/3}$ where $\alpha_f \approx 1/137$ is the fine-structure constant \cite{fedotov2022advances,yakimenko2019prospect,ritus1970radiative,meuren2011quantum}. For values $\alpha_f \chi^{2/3} \gtrsim 1$ the theory is rendered nonperturbative. In this domain, photons, electrons and positrons can be thought to acquire an effective mass as a result of radiative corrections. Specifically, one can show that the effective mass of the electron $\Tilde{m}_e$ can be estimated to be \cite{yakimenko2019prospect}

\begin{equation} \label{eq:EffMass}
\Tilde{m}^2_e = m^2_e + \delta m^2_e = m^2_e \left(1 + 0.84 \alpha_f \chi^{2/3}\right)
\end{equation}
which implies an effective value of $\chi$ (mass enters Eq. \eqref{eq:chi} through $E_{\text{crit}}$)
\begin{equation}
\Tilde{\chi}^{2/3} = \frac{\chi^{2/3}}{1 + 0.84 \alpha_f \chi^{2/3}} .
\end{equation}
To benchmark this effect and measure its extent one can consider the value of $0.84$ as a model parameter $\theta$ to be determined based on experiments:
\begin{equation} \label{eq:EffChi}
\Tilde{\chi}^{2/3} = \frac{\chi^{2/3}}{1 + \theta \alpha_f \chi^{2/3}} .
\end{equation}
Replacement of effective quantities $\Tilde{m}_e, \Tilde{\chi}$ affects the rate of photon emission and pair formation. As for the former, we can write the rate as \cite{berestetskii1982quantum,baier1967quantum}:

\begin{equation} \label{eq:nonLinComptonRate}
\frac{\partial I}{\partial \omega} \left(\delta, \theta \right) = \frac{\sqrt{3} \Tilde{m}_e q_e^2 c \Tilde{\chi}(1-\delta)}{2 \pi \gamma_e \hbar} \left( F_1(\zeta) + \frac{3}{2} \delta \Tilde{\chi} \zeta F_2(\zeta) \right)
\end{equation}
where $\zeta = \frac{2}{3 \Tilde{\chi}} \frac{\delta}{1 - \delta}$, $\delta = \frac{\hbar \omega}{\Tilde{m}_e c^2 \gamma_e}$ is the photon energy with frequency $\omega$ normalized to the emitting electron energy and $F_{1}(x), F_2(x)$ denote the first and second Synchrotron functions defined by

\begin{equation} \label{eq:SynchrotronFunctions}
    F_1(y) = y \int_y^{\infty} K_{5/3}(y) dy, \ F_2(y) = y K_{2/3}(y)
\end{equation}
with $K_{\nu}(y)$ being the modified Bessel function of the second kind. \\ 

Hence, one measurable property $x_{\text{obs}}$ might be the post-collision spectrum of photons. Indeed, the effect attributed to $\theta$ may be slight and the probabilistic nature of emissions become increasingly difficult to measure by the onset of electromagnetic cascades and low-energy emissions when $\chi \gg 1$. For our proof-of-principle, we disregard pair formation and center in on the process of nonlinear Compton scattering to elude this difficulty. Additionally, we neglect the energy loss of electrons in their propagation direction under the assumption $a_0/\gamma_e \ll 1$ where $a_0 = \frac{q_e E_0}{m_e \omega_L c}$ is the peak dimensionless amplitude of the laser having frequency $\omega_L$ and peak electric field $E_0$. To define the simulator, we select an elementary geometry resembling the interaction between a focused laser pulse and a counter-propagating electron bunch, both susceptible to misalignment. We accomplish this by simulating a single electron of momentum $p_z = -m_e c \gamma$ to impinge a plane wave laser pulse with electric field

\begin{equation} \label{eq:PWPulse}
    E_x(z, t) = (1-d) E_0 \sin \left(k \xi \right) \cos^2 \left(\frac{\pi \xi}{L} \right)\Pi\left( 
 \frac{\xi}{L} \right)
\end{equation}
where $\xi = z - ct$ is the moving coordinate, $k$ and $L$ are the wavenumber and pulse length of the laser respectively and $\Pi(x)$ is defined as a function equating to unity when $|x| < 1/2$ and zero otherwise. Here we introduce the latent parameter $0 \leq d \leq 1$ to express the misalignment in the experimental scheme, reducing the laser amplitude experienced by the electrons. \\

However, the unruliness of $d$ can obstruct ABC sampling. This becomes evident by comparing the spectra produced by Eq. \eqref{eq:nonLinComptonRate} with $\theta = 0, d \neq 0$ and $\theta \neq 0, d = 0$. Writing the order of estimate for Eq. \eqref{eq:EffChi} as

\begin{equation}
    \Tilde{\chi}(\theta, d) \sim (1-d) \chi_0 \left(1+\theta \alpha_f \left((1-d) \chi_0\right)^{2/3}  \right)^{-3/2}
\end{equation}

where $\chi_0 = \gamma_e \left(E_0/E_{\text{crit}}\right)$ is the peak value of $\chi$. The two cases can yield comparable values $\Tilde{\chi}(\theta\neq0, d=0) \sim \Tilde{\chi}(\theta=0, d\neq0)$ if

\begin{equation}
    d \sim 1 - \left(1+\theta \alpha_f \chi_0^{2/3}  \right)^{-3/2}.
\end{equation}
As a result, the value of $\Tilde{\chi}$ can be similar for several combinations of $\theta$ and $d$, generating similar energy spectra. Hence, any summary statistic obtained from such data can be near-identical, obscuring the effect of $d$ to that of $\theta$ or vice versa. Conclusively, the energy spectrum is not indicative enough to infer the value of $\theta$. This can be remedied by including information into $x_{\text{obs}}$ such that the effects of $\theta$ and $d$ become disentangled. If a complementary property of the emission is found such that the induced deviation of either parameter becomes uncorrelated, it is possible to disentangle their effects on $x_{\text{obs}}$. \\

We now seek such a property to be included into $x_{\text{obs}}$ and the choice of summary statistics to eliminate the latent variable $d$. To commence the discussion we remark that electrons conserve their transverse momentum within the laser field \cite{macchi2013superintense}

\begin{equation} \label{eq:TransverseMomentum}
\vec{p}_{\perp} = q_e \int \vec{E}_{\perp} dt
\end{equation}
in which $\vec{p}_{\perp}$ and $\vec{E}_{\perp}$ denote the transverse components of the electron momentum and electric field respectively. Therefore, at each instance of time, the electron propagates towards the direction that deviates from the initial direction by an angle $\alpha$:

\begin{equation} \label{eq:Alpha}
    \alpha = \arctan \left( \frac{|\vec{p}_{\perp}|}{|\vec{p}_z|} \right),
\end{equation}
where we assume that the motion remains highly relativistic. Evidently, emitted photons retain this angle and if the pulse is circularly polarized, this becomes correlated to the value of $\chi$ \cite{olofsson2022attaining}. 
Note that in the case of highly relativistic motion with $\alpha \ll 1$, the change of effective mass doesn't affect the deviation angle because it cannot change $\vec{p}_z$ due to momentum conservation (the gamma factor changes instead), while $\vec{p}_\perp$ is totally defined by the vector potential according to Eq.~\eqref{eq:TransverseMomentum}. \\

Accounting for the angular distribution of the emission leads us to redefine $x_{\text{obs}}$ as a fractional energy distribution per unit frequency $\Delta \omega$ and unit angle $\Delta \alpha$: $x_{\text{obs}}(\delta, \delta+\Delta \delta, \alpha, \alpha+\Delta \alpha)$ as a function of $\delta$ and $\alpha$. \\

\begin{figure}[ht] 
\centering
\includegraphics[width=\columnwidth]{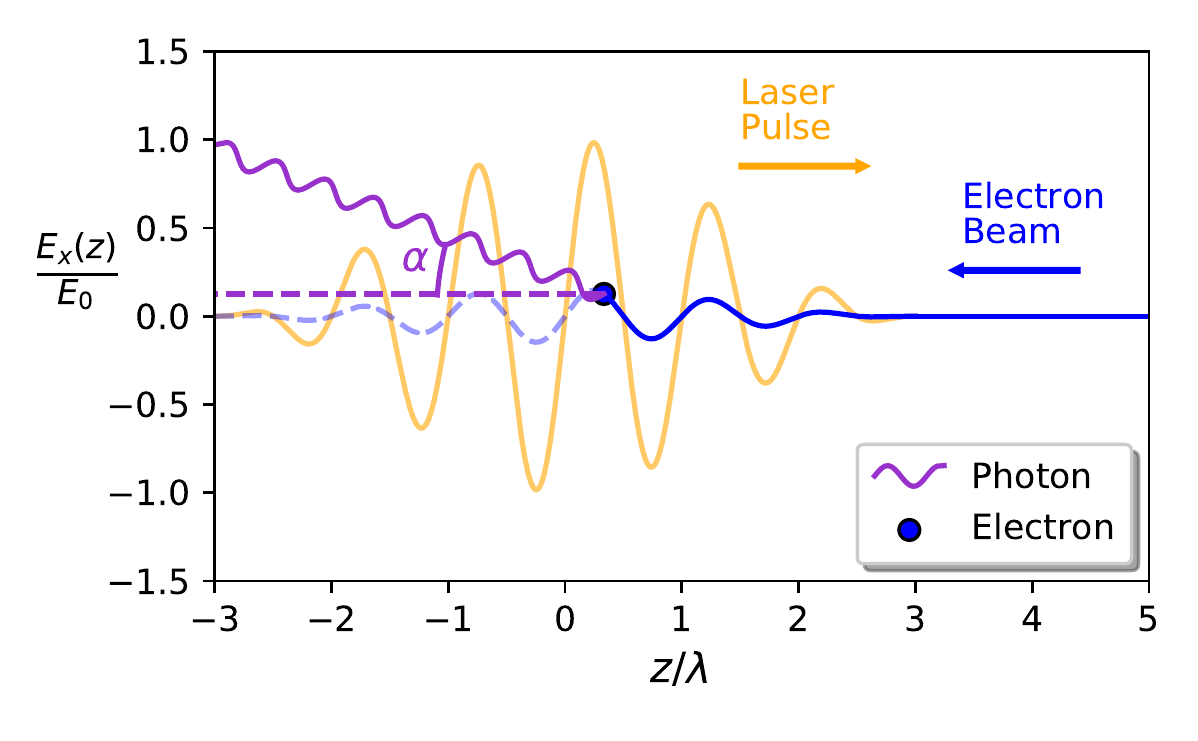}
\caption{Representation of the numerical implementation of the experiment (deviation angle is exaggerated).}
\label{fig:1Dscheme} 
\end{figure}
\begin{figure}[ht]
\centering
\includegraphics[width=\columnwidth]{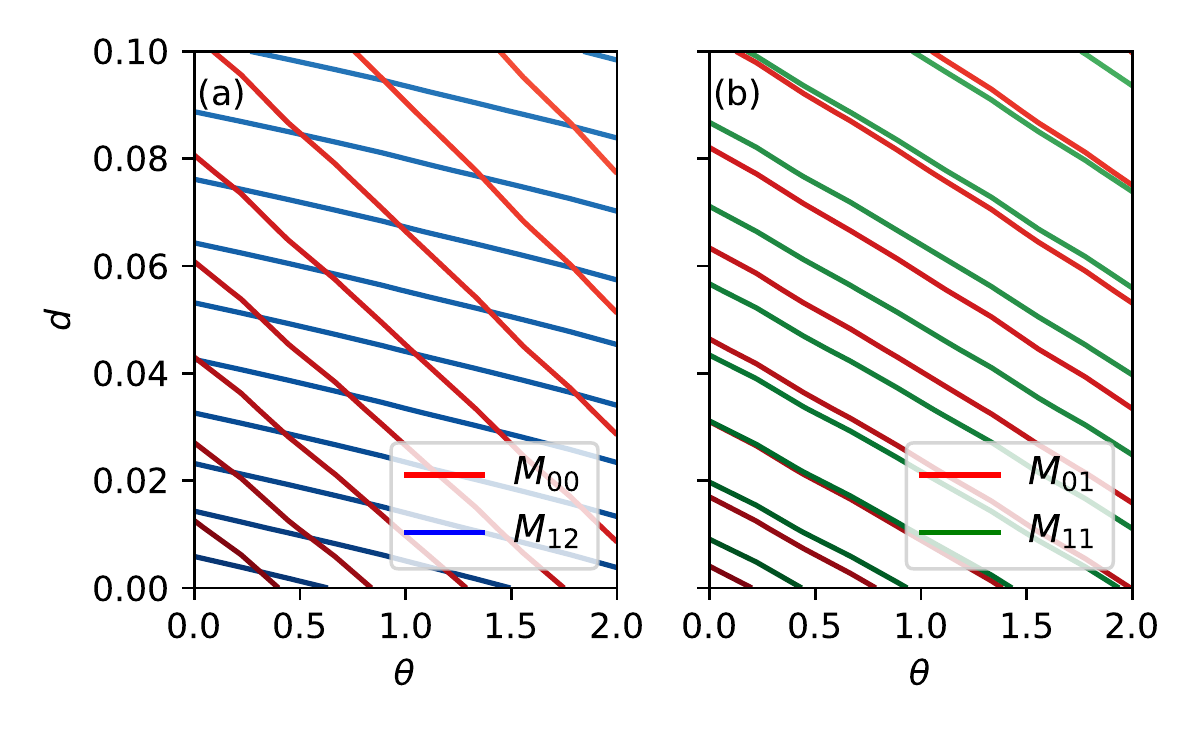}
\caption{\label{fig:Disentangle} Contours of $M_{ij}$ as a function of $\theta$ and $d$ where (a) compares $M_{00}$ and $M_{12}$, (b) compares $M_{01}$ and $M_{11}$.}
\end{figure}

We are now in a position to determine the summary statistics $S(x_{\text{obs}})$ necessary to eliminate $d$. Presumably, there exist several configurations that provide this as there is no prescribed way of formulating $S$. To identify some robust and simple enough option we evaluate moments of the two-dimensional data $x_{\text{obs}}$ to order $i$ and $j$:

\begin{equation} \label{eq:SMoments}
    M_{ij} = \int \int x_{\text{obs}}(\delta, \delta+\Delta \delta, \alpha, \alpha+\Delta \alpha) \delta^i \alpha^j d\delta d\alpha.
\end{equation}
Now, let us try to select a set of moments such that any combination $(\theta, d)$ maps to a presumably unique value of this set. Fig. \ref{fig:Disentangle} illustrates contours of four distinct moments $M_{ij}$ in the space of $\theta$ and $d$. The set of moments in Fig. \ref{fig:Disentangle}~(a) is a practical choice as the contours are not parallel anywhere, suggesting a unique pair for every $\theta$ and $d$. In contrast, Fig. \ref{fig:Disentangle}~(b) depicts a scenario when the contours become parallel at several points in the parameter space, meaning that the values of the plotted moments do not unambiguously indicate a single pair of $\theta$ and $d$. We conclude that selecting $S(x_{\text{obs}}) = \left(M_{00}, M_{12}\right)$ is a valid choice for ABC sampling.

\section{Analysis} \label{sec:analysis} 
In our simulations, the plane wave pulse is designated by a wavelength of $\lambda = \SI{0.8}{\micro\meter}$, pulse length $L = 6 \lambda$ and peak amplitude $a_0 = 100$ (excluding the factor of $(1-d)$). Electrons are assigned an initial energy of $170 ~ \si{\giga\electronvolt}$ ($\gamma_e \sim 10^5$) situated a distance $z_{\text{s}} = 5 \lambda$ from the origin (the numerical layout can be seen in Fig. \ref{fig:1Dscheme}). Both electron and pulse are allowed to counter propagate for $N$ time steps $\Delta t = \frac{\left(L + z_{\text{s}}/2 \right)c^{-1}}{N}$. Here, $x(\delta, \delta+\Delta \delta, \alpha, \alpha+\Delta \alpha)$ is discretized by a $100 \times 100$ grid of cells $x(m \Delta \delta, n \Delta \alpha)$ each with size $\Delta \delta \times \Delta \alpha$ and $m,n = 0,1,2,...,99$. At each time step $q$, Eqs. \eqref{eq:TransverseMomentum} and \eqref{eq:Alpha} are used to estimate $n \approx \alpha / \Delta \alpha$. Then, for each $m$ we accumulate 

\begin{equation} \label{eq:Dmn}
x_{q \Delta t} = x_{(q-1) \Delta t} + \Delta \alpha \Delta \omega \Delta t \frac{\partial I}{\partial \omega}\left(m\Delta \delta, \theta \right)
\end{equation}
where we have suppressed the arguments of $x$ for readability and subscripts denote the time step. For our proof-of-principle we perform blind tests of $x = M(\theta, d)$ against an "experiment" $x_{\text{obs}} = M(\theta^{true}=0.84, d)$ which serves as a ground truth. Here, the $\theta$ value is fixed to $\theta^{true}$ but the latent variable $d$ varies randomly between experiments. \\

Turning to the prerequisites for ABC sampling, we adopt the following priors over $\theta$ and $d$

\begin{equation}
    p(\theta) = \mathcal{U}(0, 150), \ p(d) = \mathcal{U}(0, 0.1)
\end{equation}
where $\mathcal{U}(a, b)$ denote the uniform distribution with lower and upper bounds $a$ and $b$ respectively. Though there is no prior knowledge apart from $\theta \geq 0$ and $0 \leq d \leq 1$ we argue that the given simulation parameters yield $\chi_0 \approx 100$ and so setting $\theta = 150$ would then drive the value of $\Tilde{\chi}$ below one, approaching a classical description. As for $d$, one could construct a prior from empirical values obtained in a real experiment. Lacking this option, we assume that the amplitude can vary at most by $10 \%$. \\

During sampling, the following distance is calculated to discriminate between observations

\begin{equation}
    ||S(x) - S(x_{\text{obs}})|| = \sqrt{d^2_{00} + d^2_{12}}
\end{equation}
where $d_{ij} = |1 - \frac{M^{\text{sim}}_{ij}}{M_{ij}}|$ (not to be confused with the latent parameter) in which the superscript label moments evaluated from simulations $x = M(\theta, d)$. A uniform kernel $K_{\epsilon}(\cdot) = \Pi(\cdot)$ is chosen with threshold $\epsilon = 0.1$ derived from the requirement to accept $N_{\theta} = 1600$ samples over the course of $\approx 50$ sampling hours. For every $50$:th proposal $\theta^*$ we generate new observed data $x_{\text{obs}}$ as to not bias the result toward the existing value of $d^* \sim p(d)$. \\ 

\begin{figure}[ht]
\centering
\includegraphics[width=\columnwidth]{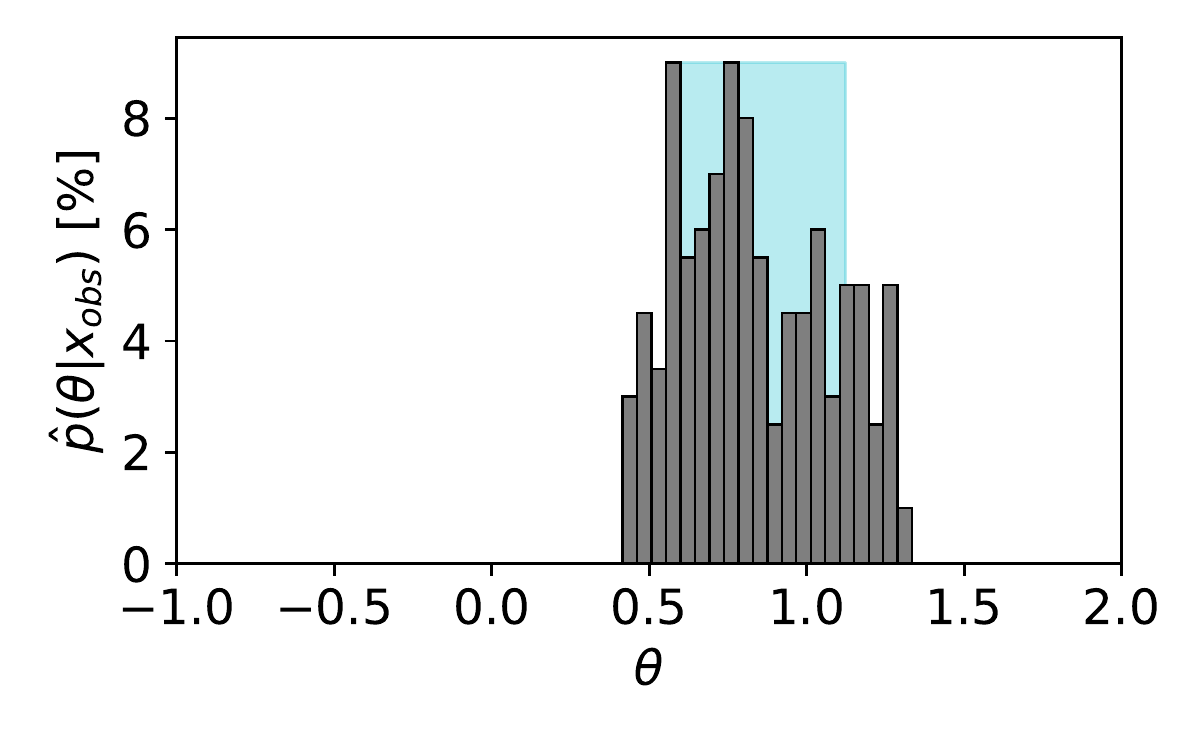}
\caption{\label{fig:Posterior} Approximate posterior obtained with $1600$ accepted samples using ABC sampling where the shaded region indicate the $68$ \% credible interval.}
\end{figure}

In Fig. \ref{fig:Posterior} we present the 
result of sampling the posterior based on the described ABC routine applied to the simulated outcome of a single collision experiment with unknown value of $d$. The fact that the accepted samples are distributed around the actually selected value of $\theta^{true} = 0.84$ indicates the claimed capability of the method. To achieve narrower distribution and reduce the credible interval of the distribution, one can process a number of experiments through Alg. \ref{alg:ABCsampling}. That is, the next experiment adapts a prior based on the inference from the previous one.

\section{Conclusions} \label{sec:Conclusion}
We have considered prospects for an experiment capable of inferring a parameter $\theta$ that signify deviations from nonlinear Compton scattering via the notion of effective mass in the regime $\chi \gg 1$. The results propel the strategies necessary to incorporate ABC sampling in analogous experiments, scalable to the inclusion of several parameters $\theta$ and $z$ accounting for alternative nonperturbative effects. An improved implementation of the interaction will be needed for designing future experiments. This can be done by e.g. simulating a realistically focused laser pulse, devising more comprehensive description via latent parameters and accounting for electromagnetic cascades. Carrying it out might pose an increased computational load as well as affect the sampling efficiency of ABC. Nonetheless, its convergence can be accelerated by further investigating additional summary statistics, non-uniform kernels and the use of machine learning to suggest better proposals. Additionally, the use of high-performance computing to recruit many ABC samplers in parallel can alleviate both impairments.

\begin{acknowledgments}
The authors acknowledge support from the Swedish Research Council (Grants No. 2017-05148 and No. 2019-02376). The computations were enabled by resources provided by the National Academic Infrastructure for Supercomputing in Sweden (NAISS) at Tetralith  partially funded by the Swedish Research Council through grant agreement no. 2022-06725. The authors would like to thank Tom Blackburn for useful discussions.
\end{acknowledgments}


\bibliography{bib_template_APS}

\providecommand{\noopsort}[1]{}\providecommand{\singleletter}[1]{#1}%
\begin{thebibliography}{26}%
\makeatletter
\providecommand \@ifxundefined [1]{%
 \@ifx{#1\undefined}
}%
\providecommand \@ifnum [1]{%
 \ifnum #1\expandafter \@firstoftwo
 \else \expandafter \@secondoftwo
 \fi
}%
\providecommand \@ifx [1]{%
 \ifx #1\expandafter \@firstoftwo
 \else \expandafter \@secondoftwo
 \fi
}%
\providecommand \natexlab [1]{#1}%
\providecommand \enquote  [1]{``#1''}%
\providecommand \bibnamefont  [1]{#1}%
\providecommand \bibfnamefont [1]{#1}%
\providecommand \citenamefont [1]{#1}%
\providecommand \href@noop [0]{\@secondoftwo}%
\providecommand \href [0]{\begingroup \@sanitize@url \@href}%
\providecommand \@href[1]{\@@startlink{#1}\@@href}%
\providecommand \@@href[1]{\endgroup#1\@@endlink}%
\providecommand \@sanitize@url [0]{\catcode `\\12\catcode `\$12\catcode
  `\&12\catcode `\#12\catcode `\^12\catcode `\_12\catcode `\%12\relax}%
\providecommand \@@startlink[1]{}%
\providecommand \@@endlink[0]{}%
\providecommand \url  [0]{\begingroup\@sanitize@url \@url }%
\providecommand \@url [1]{\endgroup\@href {#1}{\urlprefix }}%
\providecommand \urlprefix  [0]{URL }%
\providecommand \Eprint [0]{\href }%
\providecommand \doibase [0]{https://doi.org/}%
\providecommand \selectlanguage [0]{\@gobble}%
\providecommand \bibinfo  [0]{\@secondoftwo}%
\providecommand \bibfield  [0]{\@secondoftwo}%
\providecommand \translation [1]{[#1]}%
\providecommand \BibitemOpen [0]{}%
\providecommand \bibitemStop [0]{}%
\providecommand \bibitemNoStop [0]{.\EOS\space}%
\providecommand \EOS [0]{\spacefactor3000\relax}%
\providecommand \BibitemShut  [1]{\csname bibitem#1\endcsname}%
\let\auto@bib@innerbib\@empty
\bibitem [{\citenamefont {Poder}\ \emph {et~al.}(2018)\citenamefont {Poder},
  \citenamefont {Tamburini}, \citenamefont {Sarri}, \citenamefont {Di~Piazza},
  \citenamefont {Kuschel}, \citenamefont {Baird}, \citenamefont {Behm},
  \citenamefont {Bohlen}, \citenamefont {Cole}, \citenamefont {Corvan},
  \citenamefont {Duff}, \citenamefont {Gerstmayr}, \citenamefont {Keitel},
  \citenamefont {Krushelnick}, \citenamefont {Mangles}, \citenamefont
  {McKenna}, \citenamefont {Murphy}, \citenamefont {Najmudin}, \citenamefont
  {Ridgers}, \citenamefont {Samarin}, \citenamefont {Symes}, \citenamefont
  {Thomas}, \citenamefont {Warwick},\ and\ \citenamefont
  {Zepf}}]{poder2018experimental}%
  \BibitemOpen
  \bibfield  {author} {\bibinfo {author} {\bibfnamefont {K.}~\bibnamefont
  {Poder}}, \bibinfo {author} {\bibfnamefont {M.}~\bibnamefont {Tamburini}},
  \bibinfo {author} {\bibfnamefont {G.}~\bibnamefont {Sarri}}, \bibinfo
  {author} {\bibfnamefont {A.}~\bibnamefont {Di~Piazza}}, \bibinfo {author}
  {\bibfnamefont {S.}~\bibnamefont {Kuschel}}, \bibinfo {author} {\bibfnamefont
  {C.~D.}\ \bibnamefont {Baird}}, \bibinfo {author} {\bibfnamefont
  {K.}~\bibnamefont {Behm}}, \bibinfo {author} {\bibfnamefont {S.}~\bibnamefont
  {Bohlen}}, \bibinfo {author} {\bibfnamefont {J.~M.}\ \bibnamefont {Cole}},
  \bibinfo {author} {\bibfnamefont {D.~J.}\ \bibnamefont {Corvan}}, \bibinfo
  {author} {\bibfnamefont {M.}~\bibnamefont {Duff}}, \bibinfo {author}
  {\bibfnamefont {E.}~\bibnamefont {Gerstmayr}}, \bibinfo {author}
  {\bibfnamefont {C.~H.}\ \bibnamefont {Keitel}}, \bibinfo {author}
  {\bibfnamefont {K.}~\bibnamefont {Krushelnick}}, \bibinfo {author}
  {\bibfnamefont {S.~P.~D.}\ \bibnamefont {Mangles}}, \bibinfo {author}
  {\bibfnamefont {P.}~\bibnamefont {McKenna}}, \bibinfo {author} {\bibfnamefont
  {C.~D.}\ \bibnamefont {Murphy}}, \bibinfo {author} {\bibfnamefont
  {Z.}~\bibnamefont {Najmudin}}, \bibinfo {author} {\bibfnamefont {C.~P.}\
  \bibnamefont {Ridgers}}, \bibinfo {author} {\bibfnamefont {G.~M.}\
  \bibnamefont {Samarin}}, \bibinfo {author} {\bibfnamefont {D.~R.}\
  \bibnamefont {Symes}}, \bibinfo {author} {\bibfnamefont {A.~G.~R.}\
  \bibnamefont {Thomas}}, \bibinfo {author} {\bibfnamefont {J.}~\bibnamefont
  {Warwick}},\ and\ \bibinfo {author} {\bibfnamefont {M.}~\bibnamefont
  {Zepf}},\ }\bibfield  {title} {\bibinfo {title} {Experimental signatures of
  the quantum nature of radiation reaction in the field of an ultraintense
  laser},\ }\href {https://doi.org/10.1103/PhysRevX.8.031004} {\bibfield
  {journal} {\bibinfo  {journal} {Phys. Rev. X}\ }\textbf {\bibinfo {volume}
  {8}},\ \bibinfo {pages} {031004} (\bibinfo {year} {2018})}\BibitemShut
  {NoStop}%
\bibitem [{\citenamefont {Cole}\ \emph {et~al.}(2018)\citenamefont {Cole},
  \citenamefont {Behm}, \citenamefont {Gerstmayr}, \citenamefont {Blackburn},
  \citenamefont {Wood}, \citenamefont {Baird}, \citenamefont {Duff},
  \citenamefont {Harvey}, \citenamefont {Ilderton}, \citenamefont {Joglekar}
  \emph {et~al.}}]{cole2018experimental}%
  \BibitemOpen
  \bibfield  {author} {\bibinfo {author} {\bibfnamefont {J.}~\bibnamefont
  {Cole}}, \bibinfo {author} {\bibfnamefont {K.}~\bibnamefont {Behm}}, \bibinfo
  {author} {\bibfnamefont {E.}~\bibnamefont {Gerstmayr}}, \bibinfo {author}
  {\bibfnamefont {T.}~\bibnamefont {Blackburn}}, \bibinfo {author}
  {\bibfnamefont {J.}~\bibnamefont {Wood}}, \bibinfo {author} {\bibfnamefont
  {C.}~\bibnamefont {Baird}}, \bibinfo {author} {\bibfnamefont {M.~J.}\
  \bibnamefont {Duff}}, \bibinfo {author} {\bibfnamefont {C.}~\bibnamefont
  {Harvey}}, \bibinfo {author} {\bibfnamefont {A.}~\bibnamefont {Ilderton}},
  \bibinfo {author} {\bibfnamefont {A.}~\bibnamefont {Joglekar}}, \emph
  {et~al.},\ }\bibfield  {title} {\bibinfo {title} {Experimental evidence of
  radiation reaction in the collision of a high-intensity laser pulse with a
  laser-wakefield accelerated electron beam},\ }\href@noop {} {\bibfield
  {journal} {\bibinfo  {journal} {Physical Review X}\ }\textbf {\bibinfo
  {volume} {8}},\ \bibinfo {pages} {011020} (\bibinfo {year}
  {2018})}\BibitemShut {NoStop}%
\bibitem [{\citenamefont {Abramowicz}\ \emph {et~al.}(2021)\citenamefont
  {Abramowicz}, \citenamefont {Acosta}, \citenamefont {Altarelli},
  \citenamefont {Assmann}, \citenamefont {Bai}, \citenamefont {Behnke},
  \citenamefont {Benhammou}, \citenamefont {Blackburn}, \citenamefont
  {Boogert}, \citenamefont {Borysov} \emph
  {et~al.}}]{abramowicz2021conceptual}%
  \BibitemOpen
  \bibfield  {author} {\bibinfo {author} {\bibfnamefont {H.}~\bibnamefont
  {Abramowicz}}, \bibinfo {author} {\bibfnamefont {U.}~\bibnamefont {Acosta}},
  \bibinfo {author} {\bibfnamefont {M.}~\bibnamefont {Altarelli}}, \bibinfo
  {author} {\bibfnamefont {R.}~\bibnamefont {Assmann}}, \bibinfo {author}
  {\bibfnamefont {Z.}~\bibnamefont {Bai}}, \bibinfo {author} {\bibfnamefont
  {T.}~\bibnamefont {Behnke}}, \bibinfo {author} {\bibfnamefont
  {Y.}~\bibnamefont {Benhammou}}, \bibinfo {author} {\bibfnamefont
  {T.}~\bibnamefont {Blackburn}}, \bibinfo {author} {\bibfnamefont
  {S.}~\bibnamefont {Boogert}}, \bibinfo {author} {\bibfnamefont
  {O.}~\bibnamefont {Borysov}}, \emph {et~al.},\ }\bibfield  {title} {\bibinfo
  {title} {Conceptual design report for the luxe experiment},\ }\href@noop {}
  {\bibfield  {journal} {\bibinfo  {journal} {The European Physical Journal
  Special Topics}\ }\textbf {\bibinfo {volume} {230}},\ \bibinfo {pages} {2445}
  (\bibinfo {year} {2021})}\BibitemShut {NoStop}%
\bibitem [{\citenamefont {Yakimenko}\ \emph
  {et~al.}(2019{\natexlab{a}})\citenamefont {Yakimenko}, \citenamefont
  {Alsberg}, \citenamefont {Bong}, \citenamefont {Bouchard}, \citenamefont
  {Clarke}, \citenamefont {Emma}, \citenamefont {Green}, \citenamefont {Hast},
  \citenamefont {Hogan}, \citenamefont {Seabury} \emph
  {et~al.}}]{yakimenko2019facet}%
  \BibitemOpen
  \bibfield  {author} {\bibinfo {author} {\bibfnamefont {V.}~\bibnamefont
  {Yakimenko}}, \bibinfo {author} {\bibfnamefont {L.}~\bibnamefont {Alsberg}},
  \bibinfo {author} {\bibfnamefont {E.}~\bibnamefont {Bong}}, \bibinfo {author}
  {\bibfnamefont {G.}~\bibnamefont {Bouchard}}, \bibinfo {author}
  {\bibfnamefont {C.}~\bibnamefont {Clarke}}, \bibinfo {author} {\bibfnamefont
  {C.}~\bibnamefont {Emma}}, \bibinfo {author} {\bibfnamefont {S.}~\bibnamefont
  {Green}}, \bibinfo {author} {\bibfnamefont {C.}~\bibnamefont {Hast}},
  \bibinfo {author} {\bibfnamefont {M.}~\bibnamefont {Hogan}}, \bibinfo
  {author} {\bibfnamefont {J.}~\bibnamefont {Seabury}}, \emph {et~al.},\
  }\bibfield  {title} {\bibinfo {title} {Facet-ii facility for advanced
  accelerator experimental tests},\ }\href@noop {} {\bibfield  {journal}
  {\bibinfo  {journal} {Physical Review Accelerators and Beams}\ }\textbf
  {\bibinfo {volume} {22}},\ \bibinfo {pages} {101301} (\bibinfo {year}
  {2019}{\natexlab{a}})}\BibitemShut {NoStop}%
\bibitem [{\citenamefont {Bula}\ \emph {et~al.}(1996)\citenamefont {Bula},
  \citenamefont {McDonald}, \citenamefont {Prebys}, \citenamefont {Bamber},
  \citenamefont {Boege}, \citenamefont {Kotseroglou}, \citenamefont
  {Melissinos}, \citenamefont {Meyerhofer}, \citenamefont {Ragg}, \citenamefont
  {Burke} \emph {et~al.}}]{bula1996observation}%
  \BibitemOpen
  \bibfield  {author} {\bibinfo {author} {\bibfnamefont {C.}~\bibnamefont
  {Bula}}, \bibinfo {author} {\bibfnamefont {K.}~\bibnamefont {McDonald}},
  \bibinfo {author} {\bibfnamefont {E.}~\bibnamefont {Prebys}}, \bibinfo
  {author} {\bibfnamefont {C.}~\bibnamefont {Bamber}}, \bibinfo {author}
  {\bibfnamefont {S.}~\bibnamefont {Boege}}, \bibinfo {author} {\bibfnamefont
  {T.}~\bibnamefont {Kotseroglou}}, \bibinfo {author} {\bibfnamefont
  {A.}~\bibnamefont {Melissinos}}, \bibinfo {author} {\bibfnamefont
  {D.}~\bibnamefont {Meyerhofer}}, \bibinfo {author} {\bibfnamefont
  {W.}~\bibnamefont {Ragg}}, \bibinfo {author} {\bibfnamefont {D.}~\bibnamefont
  {Burke}}, \emph {et~al.},\ }\bibfield  {title} {\bibinfo {title} {Observation
  of nonlinear effects in compton scattering},\ }\href@noop {} {\bibfield
  {journal} {\bibinfo  {journal} {Physical Review Letters}\ }\textbf {\bibinfo
  {volume} {76}},\ \bibinfo {pages} {3116} (\bibinfo {year}
  {1996})}\BibitemShut {NoStop}%
\bibitem [{\citenamefont {Iinuma}\ \emph {et~al.}(2005)\citenamefont {Iinuma},
  \citenamefont {Matsukado}, \citenamefont {Endo}, \citenamefont {Hashida},
  \citenamefont {Hayashi}, \citenamefont {Kohara}, \citenamefont {Matsumoto},
  \citenamefont {Nakanishi}, \citenamefont {Sakabe}, \citenamefont {Shimizu}
  \emph {et~al.}}]{iinuma2005observation}%
  \BibitemOpen
  \bibfield  {author} {\bibinfo {author} {\bibfnamefont {M.}~\bibnamefont
  {Iinuma}}, \bibinfo {author} {\bibfnamefont {K.}~\bibnamefont {Matsukado}},
  \bibinfo {author} {\bibfnamefont {I.}~\bibnamefont {Endo}}, \bibinfo {author}
  {\bibfnamefont {M.}~\bibnamefont {Hashida}}, \bibinfo {author} {\bibfnamefont
  {K.}~\bibnamefont {Hayashi}}, \bibinfo {author} {\bibfnamefont
  {A.}~\bibnamefont {Kohara}}, \bibinfo {author} {\bibfnamefont
  {F.}~\bibnamefont {Matsumoto}}, \bibinfo {author} {\bibfnamefont
  {Y.}~\bibnamefont {Nakanishi}}, \bibinfo {author} {\bibfnamefont
  {S.}~\bibnamefont {Sakabe}}, \bibinfo {author} {\bibfnamefont
  {S.}~\bibnamefont {Shimizu}}, \emph {et~al.},\ }\bibfield  {title} {\bibinfo
  {title} {Observation of second harmonics in laser--electron scattering using
  low energy electron beam},\ }\href@noop {} {\bibfield  {journal} {\bibinfo
  {journal} {Physics Letters A}\ }\textbf {\bibinfo {volume} {346}},\ \bibinfo
  {pages} {255} (\bibinfo {year} {2005})}\BibitemShut {NoStop}%
\bibitem [{\citenamefont {Kumita}\ \emph {et~al.}(2006)\citenamefont {Kumita},
  \citenamefont {Kamiya}, \citenamefont {Babzien}, \citenamefont {Ben-Zvi},
  \citenamefont {Kusche}, \citenamefont {Pavlishin}, \citenamefont
  {Pogorelsky}, \citenamefont {Siddons}, \citenamefont {Yakimenko},
  \citenamefont {Hirose} \emph {et~al.}}]{kumita2006observation}%
  \BibitemOpen
  \bibfield  {author} {\bibinfo {author} {\bibfnamefont {T.}~\bibnamefont
  {Kumita}}, \bibinfo {author} {\bibfnamefont {Y.}~\bibnamefont {Kamiya}},
  \bibinfo {author} {\bibfnamefont {M.}~\bibnamefont {Babzien}}, \bibinfo
  {author} {\bibfnamefont {I.}~\bibnamefont {Ben-Zvi}}, \bibinfo {author}
  {\bibfnamefont {K.}~\bibnamefont {Kusche}}, \bibinfo {author} {\bibfnamefont
  {I.}~\bibnamefont {Pavlishin}}, \bibinfo {author} {\bibfnamefont
  {I.}~\bibnamefont {Pogorelsky}}, \bibinfo {author} {\bibfnamefont
  {D.}~\bibnamefont {Siddons}}, \bibinfo {author} {\bibfnamefont
  {V.}~\bibnamefont {Yakimenko}}, \bibinfo {author} {\bibfnamefont
  {T.}~\bibnamefont {Hirose}}, \emph {et~al.},\ }\bibfield  {title} {\bibinfo
  {title} {Observation of the nonlinear effect in relativistic thomson
  scattering of electron and laser beams},\ }\href@noop {} {\bibfield
  {journal} {\bibinfo  {journal} {Laser physics}\ }\textbf {\bibinfo {volume}
  {16}},\ \bibinfo {pages} {267} (\bibinfo {year} {2006})}\BibitemShut
  {NoStop}%
\bibitem [{\citenamefont {Englert}\ and\ \citenamefont
  {Rinehart}(1983)}]{englert1983second}%
  \BibitemOpen
  \bibfield  {author} {\bibinfo {author} {\bibfnamefont {T.}~\bibnamefont
  {Englert}}\ and\ \bibinfo {author} {\bibfnamefont {E.}~\bibnamefont
  {Rinehart}},\ }\bibfield  {title} {\bibinfo {title} {Second-harmonic photons
  from the interaction of free electrons with intense laser radiation},\
  }\href@noop {} {\bibfield  {journal} {\bibinfo  {journal} {Physical Review
  A}\ }\textbf {\bibinfo {volume} {28}},\ \bibinfo {pages} {1539} (\bibinfo
  {year} {1983})}\BibitemShut {NoStop}%
\bibitem [{\citenamefont {Fedotov}\ \emph {et~al.}(2022)\citenamefont
  {Fedotov}, \citenamefont {Ilderton}, \citenamefont {Karbstein}, \citenamefont
  {King}, \citenamefont {Seipt}, \citenamefont {Taya},\ and\ \citenamefont
  {Torgrimsson}}]{fedotov2022advances}%
  \BibitemOpen
  \bibfield  {author} {\bibinfo {author} {\bibfnamefont {A.}~\bibnamefont
  {Fedotov}}, \bibinfo {author} {\bibfnamefont {A.}~\bibnamefont {Ilderton}},
  \bibinfo {author} {\bibfnamefont {F.}~\bibnamefont {Karbstein}}, \bibinfo
  {author} {\bibfnamefont {B.}~\bibnamefont {King}}, \bibinfo {author}
  {\bibfnamefont {D.}~\bibnamefont {Seipt}}, \bibinfo {author} {\bibfnamefont
  {H.}~\bibnamefont {Taya}},\ and\ \bibinfo {author} {\bibfnamefont
  {G.}~\bibnamefont {Torgrimsson}},\ }\bibfield  {title} {\bibinfo {title}
  {Advances in qed with intense background fields},\ }\href@noop {} {\bibfield
  {journal} {\bibinfo  {journal} {arXiv preprint arXiv:2203.00019}\ } (\bibinfo
  {year} {2022})}\BibitemShut {NoStop}%
\bibitem [{\citenamefont {Ritto}\ \emph {et~al.}(2022)\citenamefont {Ritto},
  \citenamefont {Beregi},\ and\ \citenamefont
  {Barton}}]{ritto2022reinforcement}%
  \BibitemOpen
  \bibfield  {author} {\bibinfo {author} {\bibfnamefont {T.}~\bibnamefont
  {Ritto}}, \bibinfo {author} {\bibfnamefont {S.}~\bibnamefont {Beregi}},\ and\
  \bibinfo {author} {\bibfnamefont {D.}~\bibnamefont {Barton}},\ }\bibfield
  {title} {\bibinfo {title} {Reinforcement learning and approximate bayesian
  computation for model selection and parameter calibration applied to a
  nonlinear dynamical system},\ }\href@noop {} {\bibfield  {journal} {\bibinfo
  {journal} {Mechanical Systems and Signal Processing}\ }\textbf {\bibinfo
  {volume} {181}},\ \bibinfo {pages} {109485} (\bibinfo {year}
  {2022})}\BibitemShut {NoStop}%
\bibitem [{\citenamefont {Kennedy}\ and\ \citenamefont
  {O'Hagan}(2001)}]{kennedy2001bayesian}%
  \BibitemOpen
  \bibfield  {author} {\bibinfo {author} {\bibfnamefont {M.~C.}\ \bibnamefont
  {Kennedy}}\ and\ \bibinfo {author} {\bibfnamefont {A.}~\bibnamefont
  {O'Hagan}},\ }\bibfield  {title} {\bibinfo {title} {Bayesian calibration of
  computer models},\ }\href@noop {} {\bibfield  {journal} {\bibinfo  {journal}
  {Journal of the Royal Statistical Society: Series B (Statistical
  Methodology)}\ }\textbf {\bibinfo {volume} {63}},\ \bibinfo {pages} {425}
  (\bibinfo {year} {2001})}\BibitemShut {NoStop}%
\bibitem [{\citenamefont {DeJong}\ \emph {et~al.}(1996)\citenamefont {DeJong},
  \citenamefont {Ingram},\ and\ \citenamefont {Whiteman}}]{dejong1996bayesian}%
  \BibitemOpen
  \bibfield  {author} {\bibinfo {author} {\bibfnamefont {D.~N.}\ \bibnamefont
  {DeJong}}, \bibinfo {author} {\bibfnamefont {B.~F.}\ \bibnamefont {Ingram}},\
  and\ \bibinfo {author} {\bibfnamefont {C.~H.}\ \bibnamefont {Whiteman}},\
  }\bibfield  {title} {\bibinfo {title} {A bayesian approach to calibration},\
  }\href@noop {} {\bibfield  {journal} {\bibinfo  {journal} {Journal of
  Business \& Economic Statistics}\ }\textbf {\bibinfo {volume} {14}},\
  \bibinfo {pages} {1} (\bibinfo {year} {1996})}\BibitemShut {NoStop}%
\bibitem [{\citenamefont {Brehmer}\ \emph {et~al.}(2020)\citenamefont
  {Brehmer}, \citenamefont {Louppe}, \citenamefont {Pavez},\ and\ \citenamefont
  {Cranmer}}]{brehmer2020mining}%
  \BibitemOpen
  \bibfield  {author} {\bibinfo {author} {\bibfnamefont {J.}~\bibnamefont
  {Brehmer}}, \bibinfo {author} {\bibfnamefont {G.}~\bibnamefont {Louppe}},
  \bibinfo {author} {\bibfnamefont {J.}~\bibnamefont {Pavez}},\ and\ \bibinfo
  {author} {\bibfnamefont {K.}~\bibnamefont {Cranmer}},\ }\bibfield  {title}
  {\bibinfo {title} {Mining gold from implicit models to improve
  likelihood-free inference},\ }\href@noop {} {\bibfield  {journal} {\bibinfo
  {journal} {Proceedings of the National Academy of Sciences}\ }\textbf
  {\bibinfo {volume} {117}},\ \bibinfo {pages} {5242} (\bibinfo {year}
  {2020})}\BibitemShut {NoStop}%
\bibitem [{\citenamefont {Akeret}\ \emph {et~al.}(2015)\citenamefont {Akeret},
  \citenamefont {Refregier}, \citenamefont {Amara}, \citenamefont {Seehars},\
  and\ \citenamefont {Hasner}}]{akeret2015approximate}%
  \BibitemOpen
  \bibfield  {author} {\bibinfo {author} {\bibfnamefont {J.}~\bibnamefont
  {Akeret}}, \bibinfo {author} {\bibfnamefont {A.}~\bibnamefont {Refregier}},
  \bibinfo {author} {\bibfnamefont {A.}~\bibnamefont {Amara}}, \bibinfo
  {author} {\bibfnamefont {S.}~\bibnamefont {Seehars}},\ and\ \bibinfo {author}
  {\bibfnamefont {C.}~\bibnamefont {Hasner}},\ }\bibfield  {title} {\bibinfo
  {title} {Approximate bayesian computation for forward modeling in
  cosmology},\ }\href@noop {} {\bibfield  {journal} {\bibinfo  {journal}
  {Journal of Cosmology and Astroparticle Physics}\ }\textbf {\bibinfo {volume}
  {2015}}\bibinfo  {number} { (08)},\ \bibinfo {pages} {043}}\BibitemShut
  {NoStop}%
\bibitem [{\citenamefont {Yakimenko}\ \emph
  {et~al.}(2019{\natexlab{b}})\citenamefont {Yakimenko}, \citenamefont
  {Meuren}, \citenamefont {Del~Gaudio}, \citenamefont {Baumann}, \citenamefont
  {Fedotov}, \citenamefont {Fiuza}, \citenamefont {Grismayer}, \citenamefont
  {Hogan}, \citenamefont {Pukhov}, \citenamefont {Silva} \emph
  {et~al.}}]{yakimenko2019prospect}%
  \BibitemOpen
\bibfield  {number} {  }\bibfield  {author} {\bibinfo {author} {\bibfnamefont
  {V.}~\bibnamefont {Yakimenko}}, \bibinfo {author} {\bibfnamefont
  {S.}~\bibnamefont {Meuren}}, \bibinfo {author} {\bibfnamefont
  {F.}~\bibnamefont {Del~Gaudio}}, \bibinfo {author} {\bibfnamefont
  {C.}~\bibnamefont {Baumann}}, \bibinfo {author} {\bibfnamefont
  {A.}~\bibnamefont {Fedotov}}, \bibinfo {author} {\bibfnamefont
  {F.}~\bibnamefont {Fiuza}}, \bibinfo {author} {\bibfnamefont
  {T.}~\bibnamefont {Grismayer}}, \bibinfo {author} {\bibfnamefont
  {M.}~\bibnamefont {Hogan}}, \bibinfo {author} {\bibfnamefont
  {A.}~\bibnamefont {Pukhov}}, \bibinfo {author} {\bibfnamefont
  {L.}~\bibnamefont {Silva}}, \emph {et~al.},\ }\bibfield  {title} {\bibinfo
  {title} {Prospect of studying nonperturbative qed with beam-beam
  collisions},\ }\href@noop {} {\bibfield  {journal} {\bibinfo  {journal}
  {Physical review letters}\ }\textbf {\bibinfo {volume} {122}},\ \bibinfo
  {pages} {190404} (\bibinfo {year} {2019}{\natexlab{b}})}\BibitemShut
  {NoStop}%
\bibitem [{\citenamefont {Ritus}(1970)}]{ritus1970radiative}%
  \BibitemOpen
  \bibfield  {author} {\bibinfo {author} {\bibfnamefont {V.}~\bibnamefont
  {Ritus}},\ }\bibfield  {title} {\bibinfo {title} {Radiative effects and their
  enhancement in an intense electromagnetic field},\ }\href@noop {} {\bibfield
  {journal} {\bibinfo  {journal} {Sov. Phys. JETP}\ }\textbf {\bibinfo {volume}
  {30}},\ \bibinfo {pages} {052805} (\bibinfo {year} {1970})}\BibitemShut
  {NoStop}%
\bibitem [{\citenamefont {Meuren}\ and\ \citenamefont
  {Di~Piazza}(2011)}]{meuren2011quantum}%
  \BibitemOpen
  \bibfield  {author} {\bibinfo {author} {\bibfnamefont {S.}~\bibnamefont
  {Meuren}}\ and\ \bibinfo {author} {\bibfnamefont {A.}~\bibnamefont
  {Di~Piazza}},\ }\bibfield  {title} {\bibinfo {title} {Quantum electron
  self-interaction in a strong laser field},\ }\href@noop {} {\bibfield
  {journal} {\bibinfo  {journal} {Physical review letters}\ }\textbf {\bibinfo
  {volume} {107}},\ \bibinfo {pages} {260401} (\bibinfo {year}
  {2011})}\BibitemShut {NoStop}%
\bibitem [{\citenamefont {Tokdar}\ and\ \citenamefont
  {Kass}(2010)}]{tokdar2010importance}%
  \BibitemOpen
  \bibfield  {author} {\bibinfo {author} {\bibfnamefont {S.~T.}\ \bibnamefont
  {Tokdar}}\ and\ \bibinfo {author} {\bibfnamefont {R.~E.}\ \bibnamefont
  {Kass}},\ }\bibfield  {title} {\bibinfo {title} {Importance sampling: a
  review},\ }\href@noop {} {\bibfield  {journal} {\bibinfo  {journal} {Wiley
  Interdisciplinary Reviews: Computational Statistics}\ }\textbf {\bibinfo
  {volume} {2}},\ \bibinfo {pages} {54} (\bibinfo {year} {2010})}\BibitemShut
  {NoStop}%
\bibitem [{\citenamefont {Doucet}\ \emph {et~al.}(2001)\citenamefont {Doucet},
  \citenamefont {Freitas},\ and\ \citenamefont
  {Gordon}}]{doucet2001introduction}%
  \BibitemOpen
  \bibfield  {author} {\bibinfo {author} {\bibfnamefont {A.}~\bibnamefont
  {Doucet}}, \bibinfo {author} {\bibfnamefont {N.~d.}\ \bibnamefont
  {Freitas}},\ and\ \bibinfo {author} {\bibfnamefont {N.}~\bibnamefont
  {Gordon}},\ }\bibfield  {title} {\bibinfo {title} {An introduction to
  sequential monte carlo methods},\ }in\ \href@noop {} {\emph {\bibinfo
  {booktitle} {Sequential Monte Carlo methods in practice}}}\ (\bibinfo
  {publisher} {Springer},\ \bibinfo {year} {2001})\ pp.\ \bibinfo {pages}
  {3--14}\BibitemShut {NoStop}%
\bibitem [{\citenamefont {Brooks}\ \emph {et~al.}(2011)\citenamefont {Brooks},
  \citenamefont {Gelman}, \citenamefont {Jones},\ and\ \citenamefont
  {Meng}}]{brooks2011handbook}%
  \BibitemOpen
  \bibfield  {author} {\bibinfo {author} {\bibfnamefont {S.}~\bibnamefont
  {Brooks}}, \bibinfo {author} {\bibfnamefont {A.}~\bibnamefont {Gelman}},
  \bibinfo {author} {\bibfnamefont {G.}~\bibnamefont {Jones}},\ and\ \bibinfo
  {author} {\bibfnamefont {X.-L.}\ \bibnamefont {Meng}},\ }\href@noop {} {\emph
  {\bibinfo {title} {Handbook of markov chain monte carlo}}}\ (\bibinfo
  {publisher} {CRC press},\ \bibinfo {year} {2011})\BibitemShut {NoStop}%
\bibitem [{\citenamefont {Sisson}\ \emph {et~al.}(2018)\citenamefont {Sisson},
  \citenamefont {Fan},\ and\ \citenamefont {Beaumont}}]{sisson2018handbook}%
  \BibitemOpen
  \bibfield  {author} {\bibinfo {author} {\bibfnamefont {S.~A.}\ \bibnamefont
  {Sisson}}, \bibinfo {author} {\bibfnamefont {Y.}~\bibnamefont {Fan}},\ and\
  \bibinfo {author} {\bibfnamefont {M.}~\bibnamefont {Beaumont}},\ }\href@noop
  {} {\emph {\bibinfo {title} {Handbook of approximate Bayesian computation}}}\
  (\bibinfo  {publisher} {CRC Press},\ \bibinfo {year} {2018})\BibitemShut
  {NoStop}%
\bibitem [{\citenamefont {Cranmer}\ \emph {et~al.}(2020)\citenamefont
  {Cranmer}, \citenamefont {Brehmer},\ and\ \citenamefont
  {Louppe}}]{cranmer2020frontier}%
  \BibitemOpen
  \bibfield  {author} {\bibinfo {author} {\bibfnamefont {K.}~\bibnamefont
  {Cranmer}}, \bibinfo {author} {\bibfnamefont {J.}~\bibnamefont {Brehmer}},\
  and\ \bibinfo {author} {\bibfnamefont {G.}~\bibnamefont {Louppe}},\
  }\bibfield  {title} {\bibinfo {title} {The frontier of simulation-based
  inference},\ }\href@noop {} {\bibfield  {journal} {\bibinfo  {journal}
  {Proceedings of the National Academy of Sciences}\ }\textbf {\bibinfo
  {volume} {117}},\ \bibinfo {pages} {30055} (\bibinfo {year}
  {2020})}\BibitemShut {NoStop}%
\bibitem [{\citenamefont {Berestetskii}\ \emph {et~al.}(1982)\citenamefont
  {Berestetskii}, \citenamefont {Lifshitz},\ and\ \citenamefont
  {Pitaevskii}}]{berestetskii1982quantum}%
  \BibitemOpen
  \bibfield  {author} {\bibinfo {author} {\bibfnamefont {V.~B.}\ \bibnamefont
  {Berestetskii}}, \bibinfo {author} {\bibfnamefont {E.~M.}\ \bibnamefont
  {Lifshitz}},\ and\ \bibinfo {author} {\bibfnamefont {L.~P.}\ \bibnamefont
  {Pitaevskii}},\ }\href@noop {} {\emph {\bibinfo {title} {Quantum
  Electrodynamics: Volume 4}}},\ Vol.~\bibinfo {volume} {4}\ (\bibinfo
  {publisher} {Butterworth-Heinemann},\ \bibinfo {year} {1982})\BibitemShut
  {NoStop}%
\bibitem [{\citenamefont {Baier}\ and\ \citenamefont
  {Katkov}(1967)}]{baier1967quantum}%
  \BibitemOpen
  \bibfield  {author} {\bibinfo {author} {\bibfnamefont {V.}~\bibnamefont
  {Baier}}\ and\ \bibinfo {author} {\bibfnamefont {V.}~\bibnamefont {Katkov}},\
  }\bibfield  {title} {\bibinfo {title} {Quantum effects in magnetic
  bremsstrahlung},\ }\href@noop {} {\bibfield  {journal} {\bibinfo  {journal}
  {Physics Letters A}\ }\textbf {\bibinfo {volume} {25}},\ \bibinfo {pages}
  {492} (\bibinfo {year} {1967})}\BibitemShut {NoStop}%
\bibitem [{\citenamefont {Macchi}(2013)}]{macchi2013superintense}%
  \BibitemOpen
  \bibfield  {author} {\bibinfo {author} {\bibfnamefont {A.}~\bibnamefont
  {Macchi}},\ }\href@noop {} {\emph {\bibinfo {title} {A superintense
  laser-plasma interaction theory primer}}}\ (\bibinfo  {publisher} {Springer
  Science \& Business Media},\ \bibinfo {year} {2013})\BibitemShut {NoStop}%
\bibitem [{\citenamefont {Olofsson}\ and\ \citenamefont
  {Gonoskov}(2022)}]{olofsson2022attaining}%
  \BibitemOpen
  \bibfield  {author} {\bibinfo {author} {\bibfnamefont {C.}~\bibnamefont
  {Olofsson}}\ and\ \bibinfo {author} {\bibfnamefont {A.}~\bibnamefont
  {Gonoskov}},\ }\bibfield  {title} {\bibinfo {title} {Attaining a strong-field
  qed signal at laser-electron colliders with optimized focusing},\ }\href@noop
  {} {\bibfield  {journal} {\bibinfo  {journal} {Physical Review A}\ }\textbf
  {\bibinfo {volume} {106}},\ \bibinfo {pages} {063512} (\bibinfo {year}
  {2022})}\BibitemShut {NoStop}%
\end{thebibliography}%

\end{document}